\documentclass[12pt]{article}
\usepackage{graphicx}       
\usepackage[T1]{fontenc}       
\renewcommand{\sb}{${\cal{S}}{\cal{B}}$}
\newcommand{\sm}{${\cal{S}}{\cal{M}}$}
\newcommand{\susy}{{{\cal SUSY}}}
\newcommand{\sww}{\sin^2\theta_W}
\newcommand{\mzz}{M_Z^2}
\newcommand{\mw}{M_W}
\newcommand{\mww}{M_W^2}
\newcommand{\np}{Nucl.\,Phys.\,}
\newcommand{\pl}{Phys.\,Lett.\,}
\newcommand{\pr}{Phys.\,Rev.\,}
\newcommand{\prl}{Phys.\,Rev.\,Lett.\,}
\newcommand{\prep}{Phys.\,Rep.\,}
\newcommand{\zp}{Z.\,Phys.\,}
\newcommand{\mt}{m_t}
\newcommand{\mtsq}{m_t^2}
\newcommand{\sbo}{\tilde{b}_1}
\newcommand{\msbo}{m_{\sbo}}
\renewcommand{\stop}{{\tilde{t}}}
\newcommand{\sto}{\tilde{t}_1}
\newcommand{\stt}{\tilde{t}_2}
\newcommand{\stl}{\tilde{t}_L}
\newcommand{\str}{\tilde{t}_R}
\newcommand{\cosbb}{\cos(2\beta)}
\newcommand{\tgb}{\tan\beta}
\newcommand{\sint}{\sin\theta_{\stop}}
\newcommand{\cost}{\cos\theta_{\stop}}
\newcommand{\sintsq}{\sin^2\theta_{\stop}}
\newcommand{\costsq}{\cos^2\theta_{\stop}}
\newcommand{\msto}{m_{\sto}}
\newcommand{\mstt}{m_{\stt}}
\newcommand{\mstosq}{m_{\sto}^2}
\newcommand{\msttsq}{m_{\stt}^2}
\begin{document}
\begin{titlepage}
\mbox{}\vfill
\begin{center}
{\large\bf Associated stop Higgs Production at the Linear Collider and
Extraction of the stop Parameters.}
\vskip2\baselineskip
{\bf G.~B\'elanger$^{1}$, F.~Boudjema$^{1}$, T.~Kon$^{2}$ and V.~Lafage$^{3}$}
\vskip3\baselineskip
{\it 1. Laboratoire de Physique Th\'eorique}
{\large LAPTH}
\footnote{URA 14-36 du CNRS, associ\'ee \`a l'Universit\'e de Savoie.}\\
{\it Chemin de Bellevue, B.P. 110, F-74941 Annecy-le-Vieux,
Cedex, France.}\\

{\it 2. Seikei University, Musashino, Tokyo 180-8633, Japan}\\

{\it 3. High Energy Accelerator Research Organisation, {\large KEK},}\\
{\it Tsukuba, Ibaraki 305-801, Japan}\\
\end{center}

\centerline{ {\bf Abstract} }
\baselineskip=14pt
\noindent
 {\small We calculate stop stop Higgs production at the linear collider.
 Combining the measurements from the pair production of the lightest
 stop and that of the mass of the Higgs we show how, in a
 scenario where only the lightest stop and the lightest Higgs
 were accessible, one could extract the mass of the heavier stop and
 infer some useful information on the SUSY parameters.}
\vfill
\rightline{LAPTH-700/98}
\rightline{KEK-CP-079}
\rightline{hep-ph/yymmddd}
\rightline{{\large Nov. 1998}}
\end{titlepage}
\baselineskip=18pt
\def\thequation{\thesection.\arabic{equation}}
\section{Motivation and stop mass parameters}
The elucidation of the mechanism of symmetry breaking alongside
the discovery of the Higgs and the study of its properties are the
prime motivations for the construction of future colliders. It is
well known that the standard description of the Higgs sector in
the \sm{} is unsatisfactory, while the implementation of \susy{}
provides an elegant solution to the naturalness problem. Moreover
\susy{} can be made consistent with all present
data~\cite{Langacker_Fits98} and offers as a bonus a successful
framework for the gauge couplings unification. Contrary to the
\sm, \susy{} predicts an upper bound for the lightest Higgs, within
reach of the upcoming LHC and linear colliders, and perhaps even
LEP2 and the Tevatron. One important ingredient is that the
top-stop sector plays a crucial role in that it can contribute
large radiative corrections to the tree-level mass of this
Higgs~\cite{Rchiggsmass_oneloop,RChiggsmass_Ellis,
CarenaWagner_Higgs_Approx1,Rchiggsmass_CWP,higgsmass_twoloop_exact}.
Studying the top-stop connection to the Higgs and electroweak symmetry
breaking, \sb, is also important for a number of reasons. In many
\susy{} scenarios, like the popular mSUGRA~\cite{mSUGRA}, a heavy
top can, very nicely, triggers \sb. Recently there has been a
renewed interest in the possibility of having a light stop, which
in many scenarios can be the lightest scalar beside the Higgs.
With strong couplings to the Higgs a light stop can make
electroweak baryogenesis work~\cite{Stop_Baryogenesis}. In another
context, it has been pointed out that if the mixing in the stop
sector is quite large together with the presence of a lightest
stop having a mass of the order of the top mass or less, it may be
impossible to detect the lightest Higgs through the classic two
photon decay at the LHC~\cite{AbdelStop_Hgg_Loops}. Therefore,
there is clearly ample motivation for the direct study of the
stop-Higgs coupling. The latter should be considered as important
as the study of the $tth$ vertex through $e^+e^-\to tth$~\cite{eetth}.
Because the light stops that we will be considering will decay into
hadrons, we feel that for the precision measurements of the couplings,
a linear collider%
~\cite{Desy_PhysRep,DESY-TESLA,NLCZDR,eeInternationalWorkshop},
especially with a high luminosity as is currently
planned~\cite{DESY-TESLA}, is best suited. Higgs radiation off light
stops at the LHC has been studied in~\cite{stophiggs_LHC} but not from
the perspective of the extraction of the \susy{} parameters.

 Our aim in the present paper is not only to give the cross section for
$\sto\sto h$ production at a moderate energy $e^+e^-$ but also to
inquire what we may learn from such a measurement. For instance
one should ask which \susy{} parameters one can hope to extract,
especially when this process is combined with other measurements
that will, for sure, be made at the same machine. This includes
$\sto\sto$ production and a prior determination of the Higgs
mass. To further strengthen our motivations we will be considering
a scenario where besides the lightest Higgs, the lightest stop is
the next-to-lightest SUSY particle. If this scenario occurs then
within a R-parity conserving model, beside the lightest neutralino
acting as the LSP, one may only observe the lightest stop, $\sto$,
and the Higgs in a first stage $e^+e^-$ linear collider. Although
this may look meagre from the perspective of discovery one should
inquire whether we can exploit the few cross sections and
observables to extract some of the basic parameters of the model
and hopefully infer some information on those particles which are
not directly accessible. The purpose of this analysis is to show
how this could be done by exploiting both indirect effects present
in the radiative corrections to the Higgs mass as well as a
subdominant cross section, $e^+e^-\to\sto\sto h$ which a high
luminosity $e^+e^-$ with a clean environment allows.

\section{Stop parameters and $e^+e^-\to\sto\sto$}
To discuss the stop sector and define our conventions, we turn to the weak
eigenstate basis where the mass matrix in the $\stop_L$, $\stop_R$
involves the the SUSY soft-breaking masses: the common SU(2)
mass $\tilde{m}_{\tilde{Q}_3}$ and the U(1) mass
$\tilde{m}_{\tilde{U}_{3R}}$, beside the mixing,
$\tilde{m}_{\tilde{t}_{LR}}^2$
\begin{eqnarray}
m_{\tilde{t}_L}^2&=&\tilde{m}_{\tilde{Q}_3}^2 + m_t^2 +
\frac{1}{2} M_Z^2 \left(1- \frac{4}{3} \sww\right) \cos(2\beta)
\\ m_{\tilde{t}_R}^2&=&\tilde{m}_{\tilde{U}_{3R}}^2 + m_t^2 +
\frac{2}{3} M_Z^2 \sin^2\theta_W \cosbb \nonumber\\
m_{\tilde{t}_{LR}}^2&=&-m_t \left(A_t +\frac{\mu}{\tgb}\right)
\end{eqnarray}

One sees that apart from the soft SUSY-breaking parameters:
$\tilde{m}_{\tilde{Q}_3}$, $\tilde{m}_{\tilde{U}_{3R}}$ and the
tri-linear top term ($A_t$), there appears also the ubiquitous
$\tgb$ and the higgsino mass term $\mu$. In principle $\tgb$ and
$\mu$ could be reconstructed from a study of chargino and heavy
neutralino cross sections~\cite{eecharginos,JapanSusy,SusyTests_ee}
but in the scenario that we are investigating with only a light
Higgs and a light stop beside the LSP\footnote{This will then be
mostly a bino.}, these may not be accessible\footnote{Of course,
by the time the LC is running one may have some useful information
on the \susy{} parameters from the LHC, for instance.}. However, non
observation of an unstable chargino/neutralino would mean that
$|\mu|$ and $|M_2|$ (the gaugino mass parameter) are large. We
will take this constraint into account.
The stop mass eigenstates are defined through the mixing angle
$\theta_{\tilde{t}}$, with the lightest stop, $\sto$,
\begin{equation}
\sto=\cost\; \stl+ \sint\; \str{}
\end{equation}
In our case, since we are aiming at reconstructing the physical
masses, we find it useful to express the mixing angle as:
\begin{eqnarray}
\sin(2 \theta_{\stop})= \frac{2 \;
m_{\tilde{t}_{LR}}^2}{\mstosq-\msttsq}
\end{eqnarray}

The tree-level amplitude for $\sto$ pair production with a left-handed
(right-handed) electron ${{\cal M}}_{L (R)}$ may be written
as~\cite{eestopstop_polar_Nojiri,Stops_Porod}:
\begin{eqnarray}
\label{thetaf}
{{\cal M}}_L&=&{{\cal M}}_0
\biggl(\frac{2}{3} + \frac{1}{s_W^2 c_W^2}\left(\frac{1}{2}-s_W^2\right)
\left(\frac{1}{2}\cos^2\theta_{\stop}\;-\;\frac{2}{3} s_W^2\right)
\frac{s}{s-M_Z^2}\biggr)\nonumber\\
{{\cal M}}_R&=&{{\cal M}}_0
\biggl(\frac{2}{3} - \frac{1}{c_W^2}
\left(\frac{1}{2}\cos^2\theta_{\stop}\;-\;\frac{2}{3} s_W^2\right)
\frac{s}{s-M_Z^2}\biggr)
\end{eqnarray}

${{\cal M}}_0$ is completely given in terms of gauge couplings.
With $\sww=1/4$ and $s\gg\mzz$, for the $\sto$ this simplifies to
\begin{eqnarray}
{{\cal M}}_L&\simeq&\frac{2}{3} {{\cal M}}_0 \biggl(\frac{2}{3}+
\costsq \biggr) \nonumber\\
{{\cal M}}_R&\simeq&\frac{2}{3}
{{\cal M}}_0 \biggl(\frac{4}{3}- \costsq \biggr)
\end{eqnarray}

We see that from the $e^+e^-\to\sto\sto$ cross section only the
$\costsq$ can be measured. Previous studies and simulations (see for
instance~\cite{Desy_PhysRep,stop_Sopczak,stop_Sopczak2}) have shown
that with a very moderate luminosity this angle could be measured at
the few per-cent level ($\simeq 3-4\%$), with the high-luminosity
envisaged by the new TESLA design (500~fb$^{-1}$),
we can do much better\footnote{We also note that for a given
$\costsq$ the precision on its measurement improves as
$1/\sqrt{{{\cal L}}}$, assuming that the experimental error is set
by the statistics. Ultimately one needs a simulation similar
to~\cite{stop_Sopczak} that takes into account the impact of the
QCD radiative corrections~\cite{eestopstop_RC,eestopstop_RC_Review}
together with ISR and exploiting the benefits of polarisation in
the environment of a high luminosity option of the LC.}. Although
this measurement of the cosine will come with a sign ambiguity we
will see that in fact, for most cases of interest, one only needs
the square of the cosine for the calculation of the $\sto\sto h$
cross section.
 The unpolarised $\sto\sto$ cross section
is (for a fixed stop mass) lowest for $\costsq=1/3$. For this
value, the $L$ and $R$ cross sections are equal. For larger
$\costsq$ production with a left-handed electron dominates.
Incidentally, as we will discuss the same beam polarisation which
makes the $\sto\sto $ largest makes the $\sto\sto h$ larger
also. From a threshold scan we can easily infer the (physical)
mass of the stop, whereas from the absolute value of the cross
section or a ratio of the left-handed/right-handed cross section
one can measure $\costsq$~\cite{eestopstop_polar_Nojiri}. In
principle, depending on the mass of the stop and its decays one
can also extract useful information and constrain some
parameters~\cite{stop_Sopczak2}. To set an order of magnitude for
the cross section, we see that one should expect cross sections of
the order of 100~fb at 500~GeV centre-of-mass energy.


\section{$e^+e^-\to\sto\sto h$ production}
\begin{figure*}[htb]
\begin{center}
\includegraphics[width=140mm]{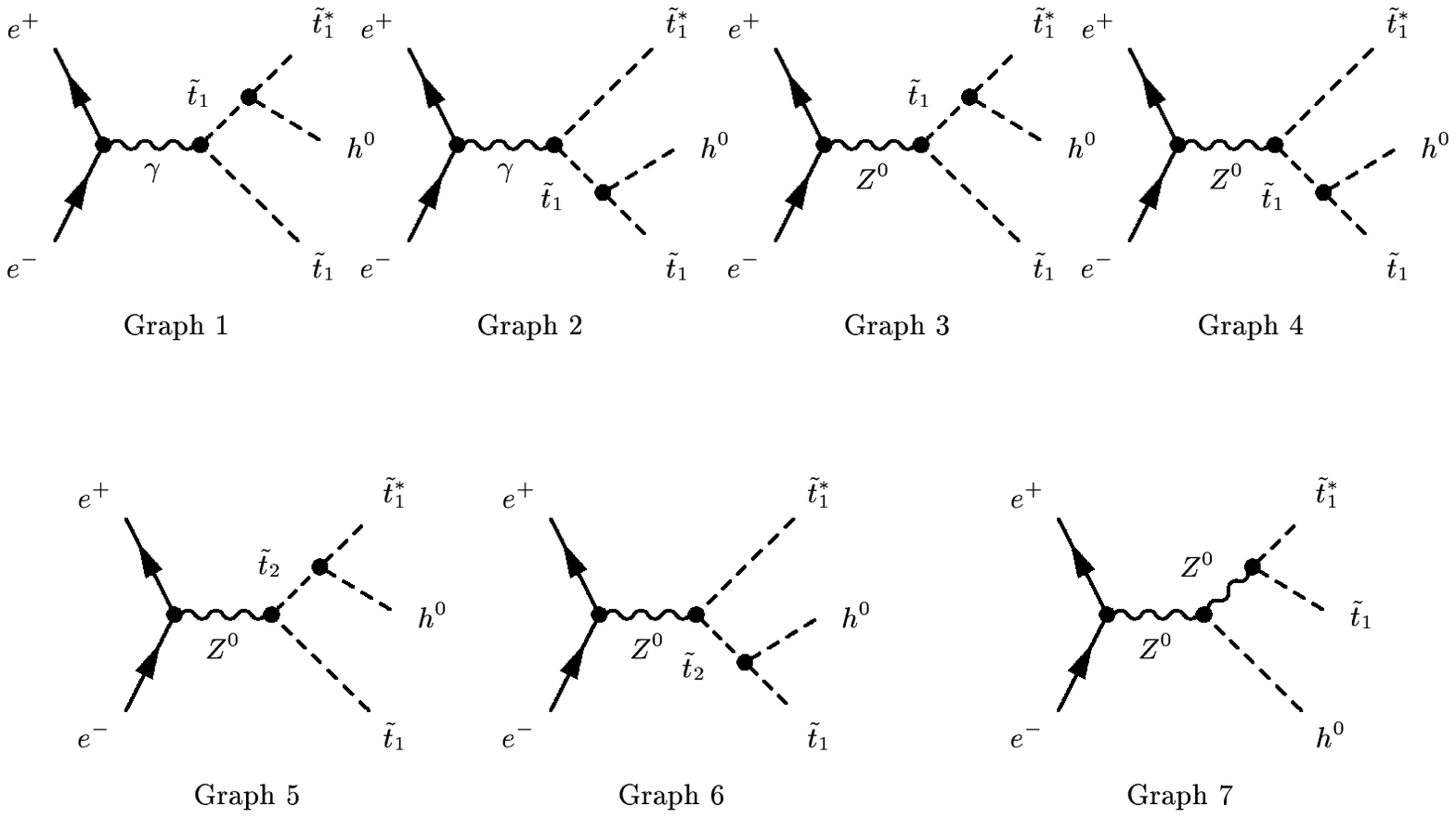}
\caption{\label{feynmangraphs}{\em Feynman graphs contributing to
$e^+e^-\to\sto\sto h$.\/}}
\end{center}
\end{figure*}
$e^+e^-\to\sto\sto h$ production proceeds via three kinds of
diagrams (See Fig.~\ref{feynmangraphs}). The most important is a
bremsstrahlung of a Higgs off $\sto$ and involves the $\sto\sto
h$ vertex (Diag.~1--4 in Fig.~\ref{feynmangraphs}). One also has
the conversion of $\stt$ to $\sto$ which involves the $\stt \sto
h$ vertex. Though this contribution turns out to be small there
are some regions in the parameter space where this contribution
can not be neglected. One last type, which we find to be
completely negligible is $hZ^*$ production with $Z^*\to\sto\sto$. 
The latter can be predicted precisely once $e^+e^-\to Zh$ 
and $e^+e^-\to\sto\sto$ have been measured, but as said this
contribution is totally negligible. The fact that the cross
section is dominated by the bremsstrahlung off the lightest stop
explains why the effect of beam polarisation on this process is
almost the same as that on the light stop pair production. Since
diagrams involving $\sto\sto h$ are dominant and could allow to
reconstruct the mass of the $\stt$ let us study this vertex in
detail.

\subsection{The $\sto\sto h$ vertex}
The stop-stop Higgs couplings, like the stop
mass matrix, emerge essentially from the F-terms in the scalar
potential (there is a residual $D$ term component $\propto\mzz$).
With the angle $\alpha$ in the Higgs mixing matrix, the $\sto\sto
h$ coupling is (we write the potential)
\begin{eqnarray}
\label{stopstophcoupling} V_{\sto\sto h}&= & -g
\frac{\mt}{\mw} \frac{\cos\alpha}{\sin\beta} \bigg(
(A^\star_t-\mu \tan\alpha) \sint\; \cost\;-\; \mt \nonumber\\
&+& \frac{\mzz}{\mt} \frac{\sin\beta}{\cos\alpha}
\sin(\alpha+\beta)\left(\left(\frac{1}{2}-\frac{2}{3}\sww\right)\costsq
+\frac{2}{3}\sww\sintsq\right)\biggr)\nonumber\\
\end{eqnarray}

The vertex does involve some important parameters which stem from
the Higgs sector, notably the angle $\alpha$. Since we will be
working in a scenario where at 500~GeV only the lightest Higgs
has been observed, we are in the the decoupling limit of large
$M_A$. Actually it has been shown that this decoupling limit is
reached for very moderate masses of $M_A$~\cite{Haber_decoupling}.
In this limit and up to radiative corrections we have,
\begin{eqnarray}
\tan\alpha\; \tan\beta=-1
\end{eqnarray}

\noindent{} This suggests to write,
\begin{eqnarray}
\tan\alpha\tan\beta=-(1+r) \qquad{\rm with} \qquad r\ll 1
\end{eqnarray}
where $r$ collects all $M_A$ dependence and other radiative
corrections which also occur in the computation of the Higgs
masses. We will return to the issue of how small $r$ is and what
parameters influence $r$, but for the moment let us observe that
for small $r$, the coupling as written in
Eq.~(\ref{stopstophcoupling}) may be cast into
\begin{eqnarray}
\label{stopstophcoupling2} V_{\sto\sto h}&=& +g R \frac{1}{\mw}
\Biggl\{\mt^2 + \sint\; \cost\left(\sint\; \cost
(\mstosq-\msttsq) -\frac{\mt \;\mu \;r}{\tgb} \right) \nonumber\\
 &+& \mzz ((2+r)\cos^2\beta- 1)
\left(\left(\frac{1}{2}-\frac{2}{3}\sww\right)\costsq+\frac{2}{3}\sww
\sintsq\right) \Biggr\} \nonumber\\ {\rm with}\;R&=&
\frac{\cos\alpha}{\sin\beta}
\quad{\rm and}\quad
R^2=\frac{1+\tan^2\beta}{1+\tan^2\beta+r^2+2r}\qquad
R\simeq 1-\frac{r}{1+\tan^2\beta}
\end{eqnarray}

\noindent{} We see that in the
limit $r \ll 1$ where $r$ is neglected, the $\sto\sto h$ very
much simplifies when written in terms of the measurable parameters
$\msto$, $\mstt$ and $\costsq$. It is important to realize that
neglecting the correction due to $r$, the coupling no longer
depends on $\mu$. Notice also that Eq.~(\ref{stopstophcoupling2})
shows that
 this
correction is reduced as $\tgb$ gets larger. However as $\tgb$
gets larger ($\tgb > 25$) effects of the sbottoms in correcting
both the vertex and the prediction of the Higgs mass start
becoming non negligible ($r$ may not be neglected in this case).
Discarding the $r$ correction altogether, we end up with a compact
formula where the only two unknowns (once $e^+e^-\to\sto\sto$
has been measured) are $\tgb$ and $\mstt$, the mass of the heavier
stop. In this approximation we get
\begin{eqnarray} \label{approxtth}
V_{\sto\sto h} &\simeq &\frac{g}{\mw}\biggl\{\sintsq\;
\costsq\;(\mstosq-\msttsq)\;+\; \mt^2\nonumber\\
&+&\mzz\cos(2\beta)
\left(\left(\frac{1}{2}-\frac{2}{3}\sww\right)\costsq+\frac{2}{3}
\sww\sintsq\right)\biggr\}
\end{eqnarray}

It should be further noticed that the explicit $\tgb$ dependence
in the vertex can be considered as subdominant compared to the
$\mtsq$ term. This remaining small D-term contribution is mildly
dependent on $\tgb$, so long as $\tgb>2$, and for values of the
stop masses and mixing such that there is a measurable $\sto\sto h$
cross section (see below). Of course, the Higgs mass does
depend quite strongly on $\tgb$ and this is where the main $\tgb$
dependence at the level of the $\sto\sto h$ cross section may be
felt. We have verified explicitly that the use of
Eq.~(\ref{approxtth}) is excellent as compared to the exact (fully
corrected) vertex. The approximation is always within $2\%$ for
all values of the stop masses and mixings that give an observable
$\sto\sto h$ cross section at a high luminosity linear collider,
as long as $\tgb<25$ for a wide scan on the sbottom masses and $\mu$. 
The vertex thus hardly shows a sensitivity to $\mu$. We also confirm
that the $\tgb$ dependence in the vertex is also hardly noticeable
and thus considering the statistical error with which the cross
section of interest will be measured the vertex will depend
essentially on the heavier stop mass (after inputing the mixing
angle and the lightest stop mass.)

Eq.~(\ref{approxtth}) makes it clear that the vertex (and hence the
cross section) will be largest for maximal mixing, 
$\sin^2 2\theta_{\stop}\sim 1$ ($\cos\theta_{\stop}\sim0.7$).
There also occurs a sharp dip in the vertex (that will also be
reflected in the cross section) when the stop contribution and that of
the top cancel each other. This occurs for values of the mixing angle
such that:
%
%
\begin{equation}
\label{dipinvertex}
\sin^2 2\theta_{\stop}\simeq\frac{4 \mt^2}{\mstt^2-\msto^2}
\end{equation}
On the other hand when the mixing is negligible, the vertex is
accounted for almost entirely by the top mass and therefore has
the same strength as the $t t h$ vertex. In this situation we
expect that if the mass of the lightest stop and that of the top
are of the same order so would the cross sections for $tth$ and
$\sto\sto h$ production. Indeed the $t t h$ vertex has strength
\begin{eqnarray}
V_{t t h}=\frac{g}{2 M_W} R \; \mt
\end{eqnarray}

Associated production of Higgs with tops has been studied
previously~\cite{eetth}. As an aside, coming back to the issue of
the large $\tgb$ and the issue of the sbottom {\em contamination\/}
in the correction to the vertex and the Higgs mass, it is worth
pointing out that this entails a not so negligible $r$. However,
the latter may be measured in the couplings of the Higgs to
fermions and bosons. In fact in this case the best way to measure
$r$ would be through a precision measurement of the Higgs decay
into $b \bar b$, which involves the coupling $R^{\prime}=R (1+r)$
which is more sensitive to $r$ than $R$ is, beside the fact that
the $h\to b\bar b$ would have by far the best statistics. In the
same vein note that $Zh$ production may not serve as a good
measure of $r$ despite its statistics since the $Z Z h$ vertex has
a (small) quadratic dependence in $r$:
\begin{eqnarray}
\label{Zzh}
\sin^2(\alpha-\beta)\simeq 1\;-\;r^2 \frac{\tgb^2}{{(1+\tgb^2)}^2}
\end{eqnarray}

Therefore if $r$ is not so small it could be extracted from
measurements of the Higgs couplings to $b\bar b$ say\footnote{A
similar conclusion was reached in~\cite{Wells_Higgs98} who
considered an expansion of the Higgs couplings in $1/M_A$ and
$1/\tgb$.}, this can then be combined with the measurements of the
cross sections that we are considering and the Higgs mass to
constrain the \susy{} parameters.

\noindent
There is another remark to make. As we pointed out above,
measurement of stop pair does not measure the absolute value of
$\cos\theta_{\stop}$ but only $\cos^2\theta_{\stop}$. It is
gratifying to see that if one neglects the small correction $\mt
\;\mu \;r/\tgb$, the $\sto\sto h$ vertex depends on the quantity
$\sin^2\theta_{\sto} \cos^2\theta_{\sto}$, that is the squares of
the mixing in the stop parameters which is exactly what is
measured from $\sto\sto$ production. This with the fact that the
$\sto\sto h$ vertex combines, at the amplitude level, with $\sto
\sto Z$ which is proportional to $\cos^2\theta_{\stop} -4/3 \sww$
(and the $\sto\sto\gamma$ which is independent of the mixing
angle) means that the sign of $\cos\theta_{\stop}$ which is not
measured in stop pair production is not critical in computing the
the $\sto$ exchange diagrams in the decoupling limit.

\subsection{Influence of the $\stt$ exchange diagrams}
 A similar conclusion holds for the amplitudes involving
$\stt$ exchange. Indeed, the $\sto\stt h$ vertex may be cast into
\begin{eqnarray}
\label{stop1stop2hcoupling2} V_{\sto\stt h}&=& +g R
\frac{1}{\mw} \Biggl\{ \frac{\cos2\theta_{\stop}}{4} \left(\sin
2\theta_{\stop}\; (\mstosq-\msttsq) -\frac{2 \mt \;\mu \;r}{\tgb}
\right) \nonumber\\ &+& \mzz \sin2\theta_{\stop}\; (\cos2\beta
+ r \cos^2 \beta)\left(\frac{2}{3} \sww-\frac{1}{4} \right)
\Biggr\}
\end{eqnarray}

Within the approximation of neglecting the $r$ terms, when
combined with the $\sto\stt Z$ vertex the full $\stt$ exchange
diagram only requires the knowledge of $\cos^2 \theta_{\stop}$.

\noindent{} Note that the term in $m_t^2$ does not appear in the
off-diagonal vertex. In the large $\mstt$ limit where the vertex
grows large the $\mstt$ dependence is off-set by that of the
propagator. The $\stt$ contribution though much smaller than that
of the dominant $\sto$ exchange diagrams may not always be
neglected. For instance take the case of a centre of mass energy
of 500~GeV, with $\mu=$400~GeV. With $\tgb=10$,
$\cos\theta_{\stop}=0.4$ and $\msto=$800~GeV we find a total cross
section of 0.78~fb, of which 90\% is accounted for by the $\sto$
diagrams alone. The $\stt$ diagrams by themselves are more than
two orders of magnitude smaller, while the $Zh$ type is a further
two orders of magnitude smaller. We find that there are points in
the parameter space where through interference the $\stt$ diagram
should be taken into account, while the $Zh$-type is invariably
always negligible. This said one should take into account that the
cross sections are rather small and even for a high luminosity of
500~fb$^{-1}$ neglecting the $\stt$ contribution hardly amounts to
more than a $2\sigma$ deviation, assuming an efficiency of $50\%$.

\subsection{The Higgs mass dependence and measurement of $\tgb$}
We have argued that for values of the stop masses and mixings
which entail a large enough stop-Higgs coupling and hence a large
cross section, the $\tgb$ dependence in the vertex is negligible.
However, the Higgs mass crucially depends on this parameter,
beside the corrections to the tree-level formula which involve the
stop parameters. In our calculation we have taken $\tgb$,
$\msto$, $\mstt$, $\cos\theta_{\stop}$ and $\mu$ as input
parameters. This allows to calculate both the vertices and the
Higgs mass in the decoupling limit and with the assumption of
vanishing sbottom contributions which we found to hold very well
so long as $\tgb$ is smaller than $\sim 25$. In this case the same
parameters that specify the $\sto\sto h$ vertex fix the Higgs
mass\footnote{This is to be expected: cutting the one-loop diagram
contributions to the Higgs mass from the top-stop sector gives the
vertices that enter the calculation of $e^+e^-\to\sto\sto h$.}.
Although in our analysis we have used numerical formulae for the
corrected Higgs mass (based on~\cite{RChiggsmass_Ellis}, but with a
running top mass to effectively incorporate the leading two-loop
corrections~\cite{CarenaWagner_Higgs_Approx1}\footnote{A more complete
analysis needs to incorporate the results of more complete
two-loop corrections~\cite{higgsmass_twoloop_exact}.}), to exhibit
the dependence of the Higgs mass on the stop parameters and help
in the discussion, it is instructive to appeal to the following
one-loop approximation (where also the mass of the top is
understood as running~\cite{Rchiggsmass_CWP}).
\begin{eqnarray}
\label{approxHiggsmass}
m_h^2&=& M_Z^2 \cos^2 (2 \beta) + \frac{3 \alpha}{4 \pi \sww}
\frac{\bar{m}_t^4}{\mww} \Biggl(\log\left(\frac{\mstosq\;
\msttsq}{m_t^4}\right) + \nonumber\\ && {\left(\sintsq\costsq
\frac{(\mstosq-\msttsq)}{\mt^2} \right)}^2 \times f
(\msttsq,\mstosq)\nonumber\\ &&
 + 2 \sintsq\costsq
\frac{(\mstosq-\msttsq)}{\mt^2}\log(\mstosq/\msttsq) \Biggr)
\nonumber\\
f(x,y)&=&2-(x+y)/(x-y)\log(x/y)
\end{eqnarray}

The latter analytical approximation shows that the combinations
that enter in the $\sto\sto h$ and the Higgs mass are the same
and most importantly that there is no $\mu$ dependence. In fact,
the numerical estimates show a very mild $\mu$ dependence (in the
limit of no sbottom contributions). Varying $|\mu|$ from 400~GeV to
1~TeV changes the Higgs mass by a few per-mil. The lower value of
400~GeV was set so that higgsinos are too heavy to be produced at a
500~GeV $e^+e^-$.

\subsection{Constraints from low Higgs masses, $\Delta\rho$, the
influence of sbottoms and CCB}
Large values of the $\sto\sto h$ vertex which lead to the largest
cross sections occur for maximal mixing with a large splitting
between the two stop physical masses. It is, however, for this
configuration that one has some strong constraints which preclude
the highest values of the cross section. For instance, one has to
be weary that imposing a lower bound on the Higgs mass, from its
non observation at LEP2 say, can restrict drastically the
$\cos\theta_{\stop}-\mstt$ parameter space. This constraint is
very much dependent on $\tgb$. Much less dependent on $\tgb$ but a
quiet powerful one, for the values of $\msto$ that we have
entertained, is the constraint coming from $\Delta
\rho$~\cite{Drhosusy}. Taking the present limit $\Delta
\rho<0.0013$ applicable to New Physics with a light
Higgs~\cite{Langacker_Fits98}, which here means essentially the
contribution from stops and sbottoms (and marginally the Higgs
sector in our decoupling scenario\footnote{For light stops in the
decoupling limit the sbottom-stop contribution when substantial
gives a positive contribution, whereas the Higgs sector
contributes a negligible negative contribution.}) generally
excludes region of the parameter space where the $\sto\sto h$ is
largest. In implementing the constraint from $\Delta\rho$ we have
taken $\msbo=$300~GeV and scanned over $\mu$ as above. In our study
the mixing in the sbottom has always been assumed to be zero.

To illustrate the effect of these constraints, let us concentrate
on the case study at 500~GeV with $\msto=$120~GeV with the
requirement that $\msbo=$300~GeV. For $\mstt=$700~GeV, the lower
bound $m_h>90$~GeV means that maximal mixing for large $\tgb$, say
$\tgb=10$, is excluded: we find that the range
$0.67<\cos\theta_{\stop}<0.73$ is not compatible with this lower bound
on the Higgs mass, whereas for $\tgb=2$ the allowed range is
$0.17 <\cos\theta_{\stop}<0.53$ and $0.85<\cos\theta_{\stop}<0.99$.
$\Delta\rho$ excludes the range $0.54<\cos\theta_{\stop}<0.94$ and
$0.98<\cos\theta_{\stop}<1$ (almost independent of $\tgb$). Putting
the two constraints, $m_h>90$~GeV, $\Delta\rho<0.0013$, together
means that the allowed range is
\begin{eqnarray}
\label{constraints}
\bullet&\tgb=10\qquad 0.~\,< \cos\theta_{\stop} <0.54
\qquad{\rm and}\qquad 0.94 < \cos\theta_{\stop} <0.98 \nonumber\\
\bullet&\tgb=2 \qquad 0.17 < \cos\theta_{\stop} <0.53
\qquad{\rm and}\qquad 0.94 < \cos\theta_{\stop} <0.98
\end{eqnarray}

For large $\tgb$ the constraint comes essentially from $\Delta
\rho$ whereas for $\tgb=2$ the constraint is from the Higgs mass.
As the mass of the heavier stop increases these ranges narrow. For
instance the above limit of 0.54 for $\tgb=10$ moves to $.48$ for
$\mstt=800$~GeV. On the other hand for moderate values of the
mixing angle, all of the above constraints constitute mild
restrictions. For instance for $\cos\theta_{\stop}=0.4$, the only
strictest condition stems from the Higgs mass. For $\tgb=10$, this
means that we should restrict ourselves to $\mstt<900$~GeV. For
$\tgb=2$ one has ``room'' in the range $450<\mstt<830$~GeV. Similar
constraints hold for the analysis at $\sqrt{s}=800$~GeV. Sticking
to $\msto=250$~GeV and $\msbo=400$~GeV we find that for
$\cos\theta_{\stop}=0.4$ the constraints are mild and are set by
the Higgs mass. For $\tgb=10(2)$ one has $\mstt<1060(1000)$~GeV,
this still allows to probe $700<\mstt<1000$~GeV. This said we will
see that taking these constraints into account still allows for
healthy cross sections especially with a high luminosity linear
collider.

To be consistent, if we stick to our scenario that no sbottom has
been observed one should take into account that one has a lower
bound on the sbottom mass, say half the center of mass energy.
However, equality of the SU(2) squarks masses means that it is not
always possible to have a light stop mass while maintaining the
lightest sbottom to be much heavier than the lightest stop. Indeed
since we are requiring the mixing angle in the sbottom sector to
be small so that any residual sbottom dependence in the Higgs and
$\sto\sto h$ is small, one of the two sbottom masses sets the
mass of the common SU(2) squark mass, up to D-terms. Therefore
when $\sto$ is mostly $\stop_L$ ($\cos\theta_{\stop} \simeq 1$)
its mass approaches the common SU(2) mass when one allows for
the top mass contribution. Therefore for $\cos\theta_{\stop}
\simeq 1$ it is generally difficult to concile a very light stop
with a much heavier sbottom which would not have been pair
produced. The constraint from the sbottom mass requires
$\cos\theta_{\stop}<0.88$ and thus reduces the possible parameter
space in Eq.~(\ref{constraints}) very little: the range
$0.94<\cos\theta_{\stop}<0.98$ will not be allowed. Nonetheless in
this situation where the left-handed sbottom may be produced, its mass
together with the measurements made in $e^+e^-\to\sto\sto$ may
be used to extract $\mstt$.

One more constraint one needs to mention. In the stop sector and
in the presence of large mixing as is the case here, one often has
to check whether the parameters do not induce colour and charge
breaking global minima (CCB). An approximate condition for this
not to happen requires that~\cite{CCBnaive}
\begin{eqnarray}
\label{CCBnaive}
A_t^2 < 3 \left(\tilde{m}_{\tilde{Q}_3}^2 +\tilde{m}_{\tilde{U}_{3R}}^2
+(M_A^2+M_Z^2)\cos^2\beta-\frac{1}{2}M_Z^2 \right)
\end{eqnarray}
This constraint can sometimes slightly reduce the parameter space
allowed by $\Delta\rho$ and the Higgs mass limit, as in
Eq.~(\ref{constraints}). However, it has been argued that this
condition may be too restrictive~\cite{CCB_Kusenko}. It was shown
that for a wide range of parameters, the global CCB minimum
becomes irrelevant on the ground that the time required to reach
the lowest energy state exceeds the present age of the universe.
Taking the tunneling rate into account the above constraint
Eq.~(\ref{CCBnaive}) is very much relaxed and can be replaced by the
mild approximate constraint~\cite{CCB_Kusenko}:
\begin{eqnarray}
\label{CCBkusenko}
A_t^2 +3\mu^2 < 
7.5 (\tilde{m}_{\tilde{Q}_3}^2 +\tilde{m}_{\tilde{U}_{3R}}^2)
\end{eqnarray}

When presenting our results we will, unless otherwise stated,
impose the limits $m_h>90$~GeV, $\Delta\rho<0.0013$ together with
the mild CCB constraint and the non observation of sbottoms at the
appropriate centre of mass energies.

\section{Results and Conclusions}
The calculation has been done by using the package
GRACE-SUSY~\cite{GraceSusy} for the automatic calculation of \susy{}
processes properly adapted to include the radiative corrections to
the Higgs mass and its couplings.

\begin{figure*}[htb]
\begin{center}
\includegraphics[width=140mm]{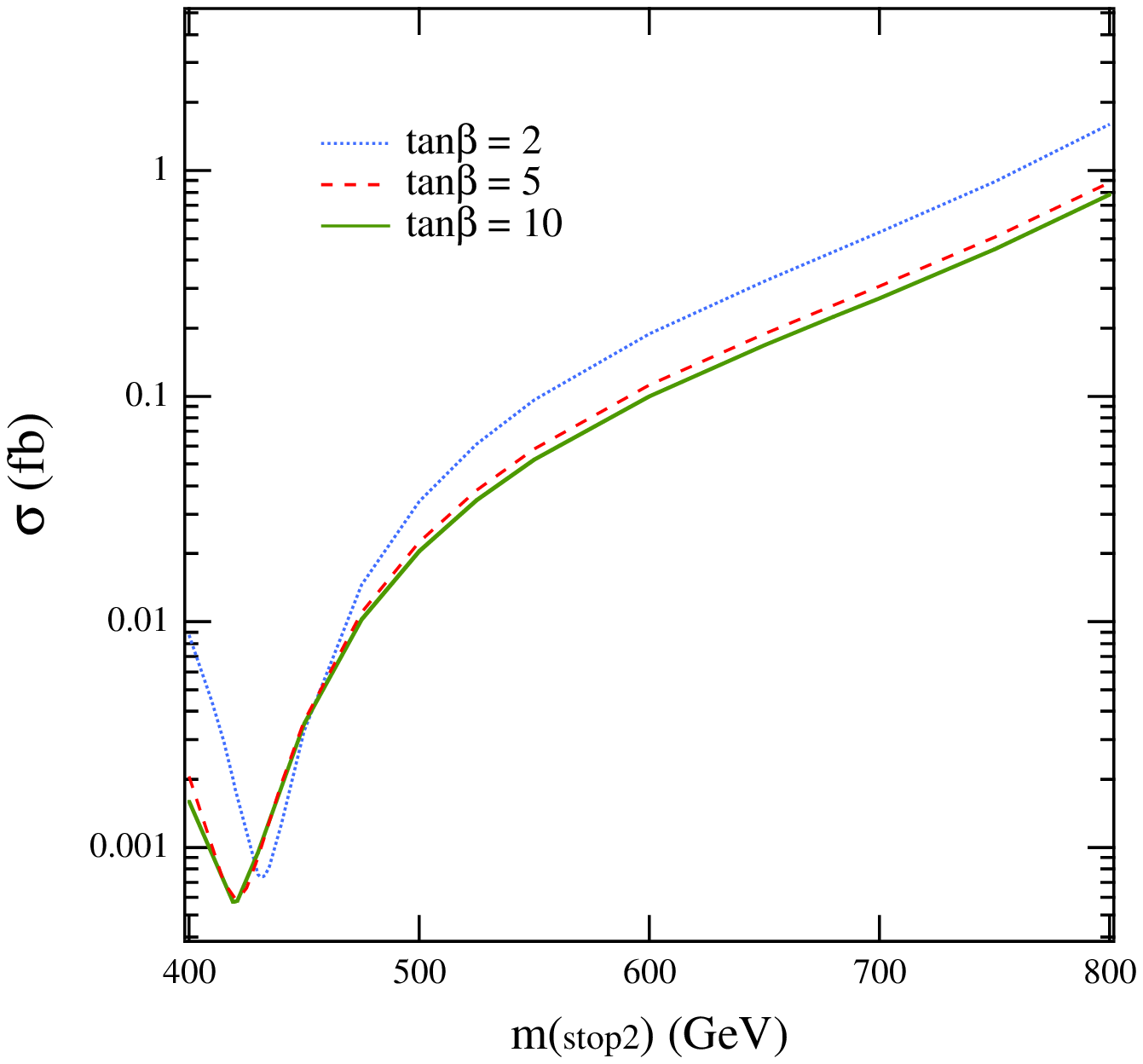}
\caption{\label{tth500tgb}{\em $\sto\sto h$ cross section at
$\sqrt{s}=500$~GeV as a function of $\mstt$ for a fixed value of
the mixing angle: $\cos\theta_{\stop}$=0.4 and $\msto=120$~GeV.
Right-handed electron polarisation is chosen.\/}}
\end{center}
\end{figure*}
\begin{figure*}[htb]
\begin{center}
\includegraphics[width=140mm,bb=0 50 380 375]{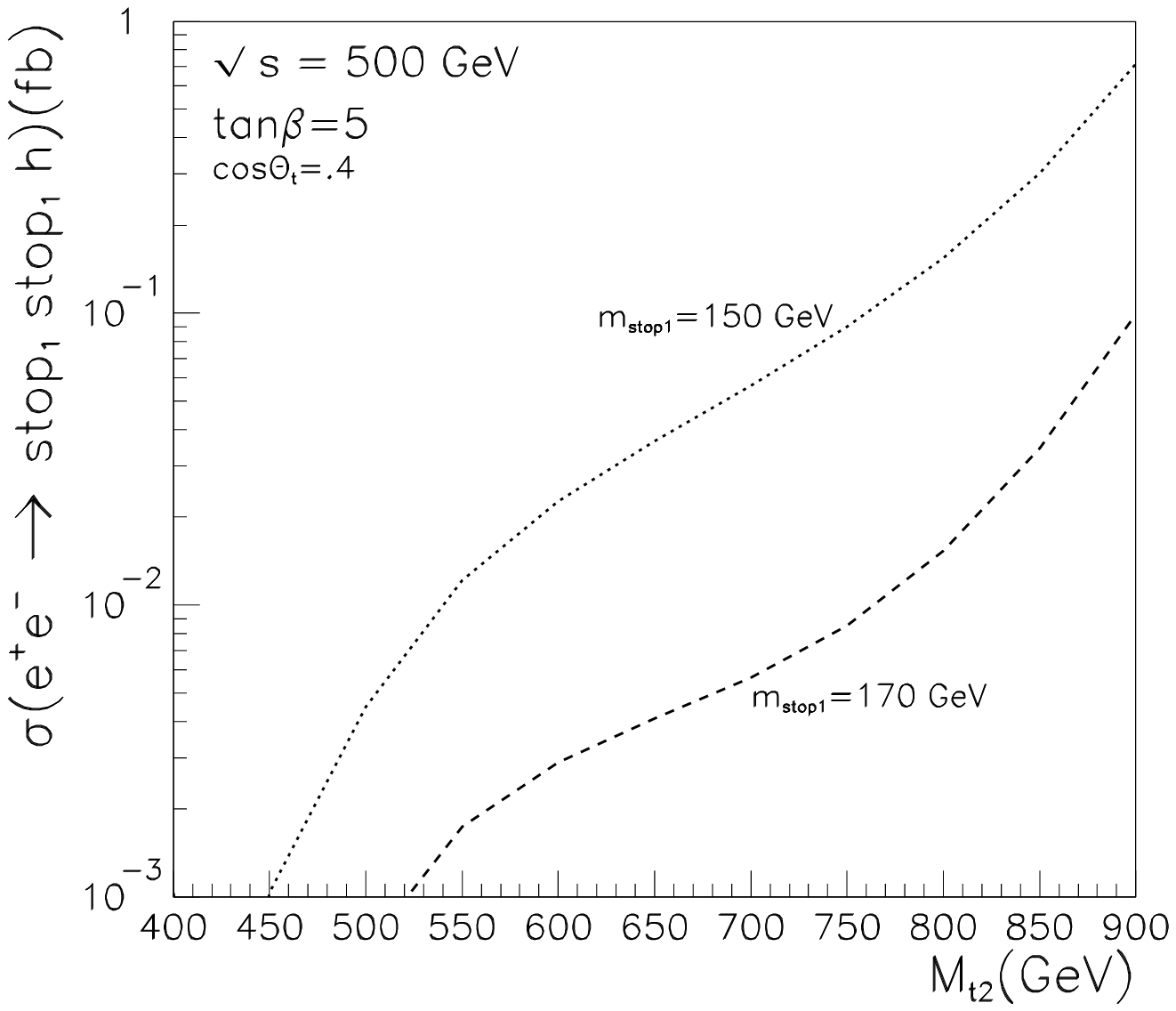}
\caption{\label{mst1_500}{\em Same as in the previous figure with
$\tgb=5$ and $\msto=150,170$~GeV.\/}}
\end{center}
\end{figure*}

\begin{figure*}[htb]
\begin{center}
\includegraphics[width=140mm,bb=86 256 465 582]{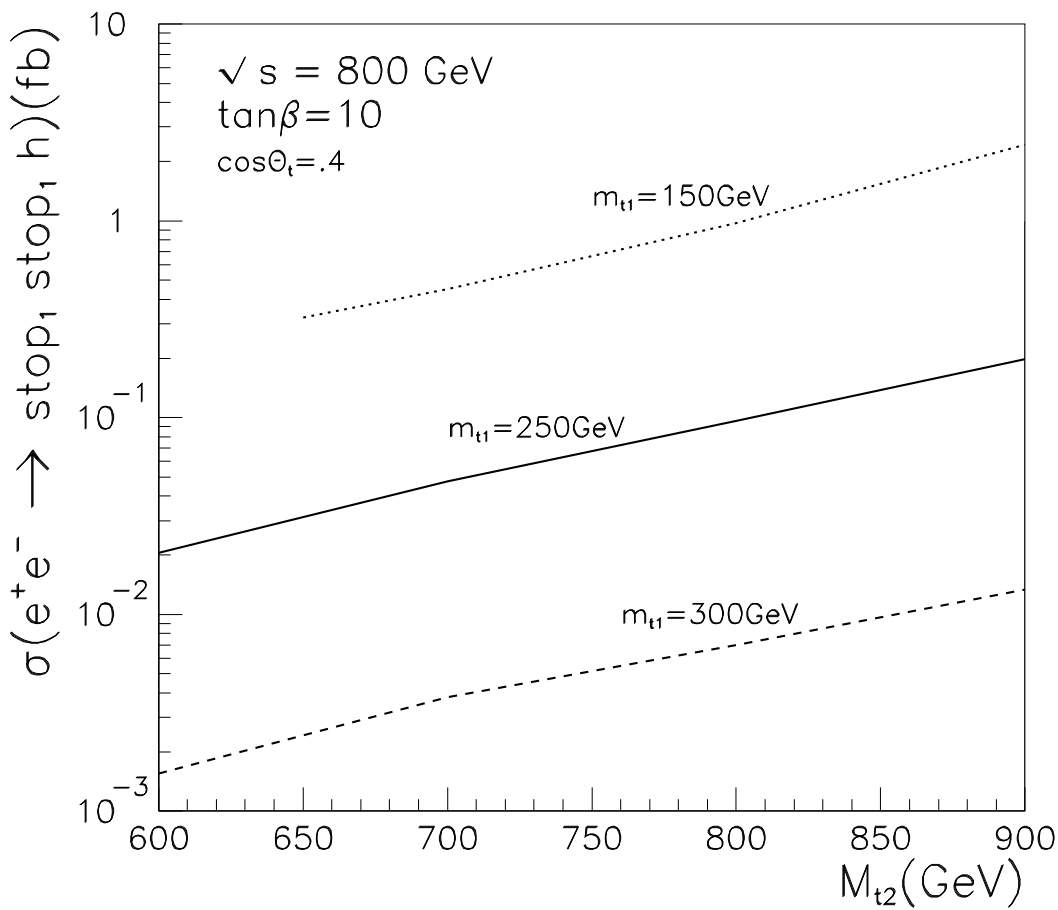}
\caption{\label{tth800}{\em Same as in the previous figure but at
$\sqrt{s}=800$~GeV for $\tgb=10$ and three representative values
of $\msto$. Note that for $\msto=150$~GeV we only consider $\stt$
masses above threshold for $\sto\stt$ production.\/}}
\end{center}
\end{figure*}

We start by describing a general feature of the cross section
which is almost independent of the centre-of-mass energy and the
input parameters: the choice of the initial polarisation. This
choice very much depends on the value of the mixing angle and is
controlled almost exclusively by the $\sto\sto$ cross section.
This is easy to understand since $\sto\sto h$ is dominated by
Higgs bremsstrahlung of $\sto$. Therefore once the stop mixing
angle has been measured one should, for the purpose of enhancing
$\sto\sto h$, choose the most appropriate polarisation which is
also beneficial for $\sto\sto$ production. For example, for
$\cos^2\theta_{\stop}< 1/3$ right-handed polarisation is best.
Let us now turn to how large cross section we should expect at a
500~GeV collider. Because of the reduced phase space we have
studied the case with $\msto=120$~GeV\footnote{The present Tevatron
limit is 122~GeV~\cite{stoplimit98} but depends strongly on the
assumed mass of the LSP neutralino and also on the mixing angle.
Taking the LSP to be heavier than 37~GeV, this limit disappears.}.
We will comment briefly on how our results change for higher masses.
Fig.~\ref{tth500tgb} shows that the expected yield depends strongly on
the mass of the heavier stop for a moderate value of the stop mixing
angle, $\cos\theta_{\stop}=0.4$ compatible with all the constraints set
in the previous section. The figure shows that for all values of
$\tgb$ there is a sharp dip in the cross section. The latter dip
corresponds to values of $\theta_{\stop}$ as given by
Eq.~(\ref{dipinvertex}) where the $\sto\sto h$ vanishes. However the
cross section picks up quickly and for $\stt$ masses larger than
$550-600$~GeV the cross section is larger than 0.1~fb, which is at the
limit of observability with a luminosity of 50~fb$^{-1}$ and should be
clearly observable with the high-luminosity TESLA option of
500~fb$^{-1}$. 
In fact, for yet larger stop masses, $\mstt=800$~GeV the cross section
is about 1~fb. 
These values depend very little on $\tgb$ whose effect is in fact
reflected in the phase space because of its influence on the Higgs
mass and therefore lower values of $\tgb$ (with all other
parameters fixed) give slightly larger cross sections since they
are associated with lower Higgs masses. Note that although the
effect occurs for a value of the cross section too small to be
observed, the location of the dip does show a slight $\tgb$
dependence. This is not surprising since this effect occurs when
the stops and top contributions in the vertex cancel each other
and hence the small $D$ term contribution may give a small
contribution. At 500~GeV the issue of phase space is crucial.
Increasing the mass of the lightest stop, the cross section drops
rather dramatically. For instance, we show in Fig.~\ref{mst1_500}
the cross section with $\msto=150$~GeV and $\msto=170$~GeV for
$\tgb=5$. We can see that compared to the case with $\msto=120$~GeV
the cross section drops by almost an order of magnitude for the
favorable value of $\mstt=800$~GeV. Increasing yet further only
slightly the mass of the lightest stop, there is no hope of observing
this process especially with a luminosity of 50~fb$^{-1}$. For higher
$\msto$ one needs to go to higher energies. We illustrate this for the
case of a 800~GeV centre-of-mass energy, see Fig.~\ref{tth800}.
At this energy one can hope to observe this process for
$\msto=250$~GeV if the mass of the heavier stop is large enough.
Alternatively, from the perspective of the extraction of the \susy{}
parameters, as the lightest stop gets heavier than the top more
and more decay channels open up, most importantly $\sto\to t
\tilde{\chi}_1^0$ which can also provide some information on the
\susy{} parameters. Note also that in all the study we conducted we
have not considered values of the $\stt$ mass such that it can be
produced in association with $\sto$ since in this case one can
have a direct measurement of the heavier stop mass. In this case
$\stt \sto$ can also trigger Higgs production through the decay of
the heavier stop to the lighter one and a Higgs, $\stt\to\sto h$,
given appropriate values of the supersymmetric parameters.

\begin{figure*}[htbp]
\begin{center}
\includegraphics[width=140mm,bb=0 50 380 375]{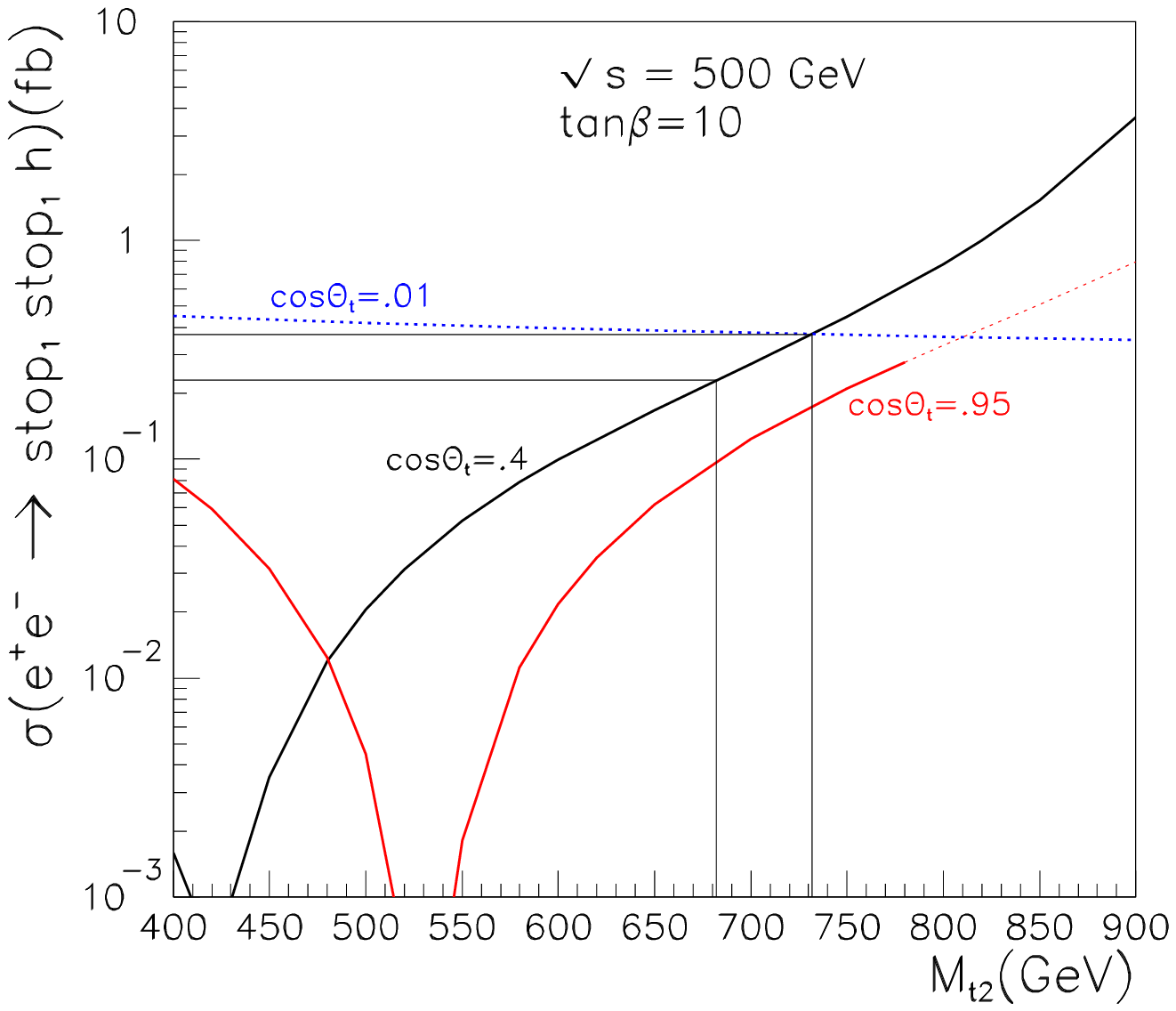}
\caption{\label{stop2extraction}{\em Extraction of the heavier
stop mass from $\sto\sto h$. The figure gives the cross section
at a centre-of-mass energy of 500~GeV as a function of the heavier
stop mass and for different mixing angles. The $3\sigma$
measurement of the mass for a mixing angle $\cos\theta_{t}$=0.4 is
also shown assuming a luminosity of 500~fb$^{-1}$ and $50\%$
detection efficiency. We have chosen the polarisation of the $e^+e^-$
such that one gets the largest cross section. For
$\cos\theta_\stop$=0.4,0.01 we took right-handed electrons whereas for
0.95 we took left-handed electrons. Note that for the latter, the
curve does not extend beyond $\mstt\simeq$780~GeV because of the
constraint from $\Delta\rho$.\/}}
\end{center}
\end{figure*}

Fig.~\ref{stop2extraction} makes clear which values of the
heaviest stop can be extracted from a measurement of $\sto\sto
h$. Note that for a no-mixing scenario the cross section although
relatively large does not allow an extraction of the heavier stop
mass. In this situation the vertex is dominated by the top mass
only and explains why the cross section is of the same order of
magnitude as the $t t h$ cross section. For large values of
$\cos\theta_{\stop}$ (0.95), there is still a sensitive $\mstt$
dependence. In this particular case however large values of
$\mstt$ ($\mstt>900$~GeV, see Fig.~\ref{stop2extraction}) are not
allowed due the $\Delta\rho$ constraint. Moreover in this
scenario SU(2) symmetry on the squark masses requires a sbottom
light enough to be produced at 500~GeV, as explained above this
allows an alternative measurement of $\mstt$. Note once again the
dramatic dip corresponding to Eq.~(\ref{dipinvertex}). With a
luminosity of 500~fb$^{-1}$ and $50\%$ detection efficiency a
$3\sigma$ measurement of the $\sto\sto h$ cross section for a
cross section of 0.3~fb restricts the mass of the heaviest stop to
$694 < m_{\stt} < 737$~GeV for a mixing angle
$\cos\theta_{\tilde{t}}=0.4$. With the requirement of 10 raw
events, a high luminosity $e^+e^-$ will allow to probe
$\mstt>550(650)$~GeV for $\cos\theta_{\tilde{t}}=0.4 (0.95)$. We also
see that for $\cos\theta_{\tilde{t}}=0.95$ $\mstt<500$~GeV can be
probed. Of course unless there is very little mixing for
$\mstt<400$~GeV the heavier stop will be discovered through $\sto
\stt$ production.

\begin{figure*}[htbp]
\begin{center}
\includegraphics[width=140mm]{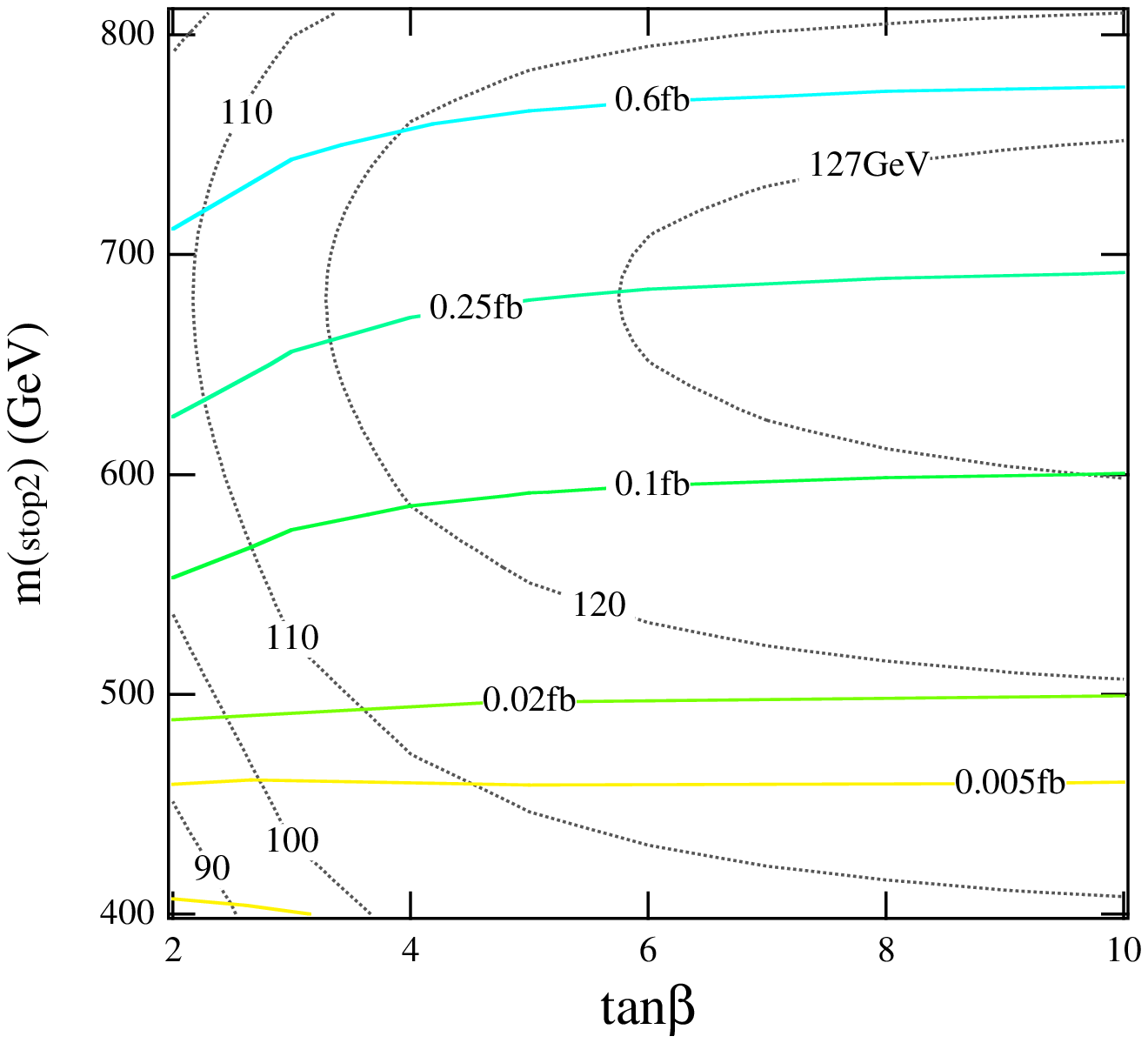}
\caption{\label{contour500}{\em Extraction of the heavier stop
mass and $\tgb$ by combining the measurement of the cross section
$\sto\sto h$ (with right-handed electrons) with that of $m_h$ at
500~GeV for $\cos\theta_{\stop}=0.4$ and $\msto=120$~GeV.\/}}
\end{center}
\end{figure*}

By combining the Higgs mass measurement and that of the $\sto\sto
h$, after having measured the stop mixing angle and the mass of
the lighter stop in $e^+e^-\to\sto\sto$, it could be possible to
extract both $\mstt$ and $\tgb$. This is shown in
Fig.~\ref{contour500} for $\cos\theta_{\tilde{t}}=0.4$. A rough
estimate based on a $3\sigma$ measurement of the $\sto\sto h$
cross section indicates that an indirect measurement of both
$\mstt$ and $\tgb$ could be achieved at the $5\%$ level. To give a
more realistic indication of how well the extraction of these
parameters can be achieved a thorough investigation is needed.
Experimentally one needs to determine how precise the mass of the
lightest stop and the mixing can be extracted from $\sto\sto$
production in a scenario where $\sto\to c \tilde{\chi}^0_1$
almost exclusively. The issue of signatures and background is
crucial. An analysis has been carried in~\cite{stop_Sopczak}. For
the parameters that we have considered we have verified that
$\sto$ decays almost exclusively into
$c\tilde{\chi}_1^0$~\cite{stoptochi0}
\footnote{For a comparison between the three-body and the two body
$c \tilde{\chi}_1^0$ decays of $\sto$ and a general discussion of
$\sto$ decays see~\cite{Stops_Porod,Porod_Wohrmann}.}. It is also
essential to include the effects of radiative corrections and
check that the latter can still allow the extraction of the mixing
angle. Indeed these corrections can introduce a few parameters
that should be disentangled from the $\cos\theta_{\stop}$
dependence. For instance, it has been shown that though the
gluonic corrections does not distort the $\cos\theta_{\tilde{t}}$
dependence, there is a slight distortion which depends on the
gluino mass~\cite{eestopstop_RC_Review}. The latter contribution
does however decrease with increasing gluino mass. Therefore, in
our scenario either the gluino is light and hence its mass
measured and in which case this could be included in the
simulation or it is heavy in which case it gives a slight
contamination to the mixing angle determination. As for the use of
$m_h$, one should stress that we have used an improved one-loop
approximation that involves only the stop parameters. A thorough
investigation should include the full two-loop corrections at
least~\cite{higgsmass_twoloop_exact}. For some region of the
parameter space, these unfortunately require the knowledge of
other \susy{} parameters. Especially in the extraction of the
parameters as illustrated in Fig.~\ref{contour500}, one should
take this into account as a theoretical error, like the neglect of
higher order corrections to the Higgs mass. In any case, the
present analysis should at least give a rough indication of how to
indirectly extract the heavy stop mass and $\tgb$, provided these
parameters yield a large enough cross section. Remark also that,
as Fig.~\ref{contour500} shows, for $\tgb>3$, $\mstt$ is given
practically exclusively by the $\sto\sto h$ cross section and
therefore one could, in a first approximation for such values of
$\tgb$, make do without a precise knowledge of $m_h$. We should also
mention that within the scenario we have pursued $e^+e^-\to\sto\sto Z$
can also provide a similar information~\cite{stopstopZ}
and thus can be used to further strengthen the conclusion of the
present analysis. In most part of the present analysis we have not
put the LHC in the picture. It is clear that some of the
parameters discussed here and perhaps more than those discussed
here could also be measured at the LHC\@. For instance a heavy stop
with a mass up to a TeV can be produced and its mass measured. For
the scenarios we have been entertaining in this paper, these
heavier stop could trigger Higgs production through their decay
into the lighter stop and a Higgs. For some of these
configurations the classic Higgs production through two-gluon with
the subsequent decay of the Higgs into two photon can be much
suppressed~\cite{AbdelStop_Hgg_Loops}. Heavier stop production can
then, among other processes, serve as an alternative to Higgs
production~\cite{Sridharandus_stophiggs_lhc}.

{\bf\large Acknowledgement}

We would like to thank the participants of the {\em 2nd Joint
ECFA/DESY Study on Physics a Linear Collider\/} for their comments
and interests. Abdel Djouadi has also informed us of a calculation
of $e^+e^-\to\sto\sto h$~\cite{eeststh_abdel}. The work of V.~L.
is supported by a JSPS Fellowship (P97215).
The work of T.~K. is supported by the Ministry of Education,
Science and Culture, Japan under Grant-in-Aid (N$^\circ$ 08640388).

\end{document}